\documentclass[12pt,a4paper]{article}

\usepackage{amsmath,epsfig}
\usepackage{amssymb,amsfonts}
\usepackage{latexsym}
\usepackage{braket}

 \newcommand{\arXiv}[1]{\href{http://www.arXiv.org/abs/#1}{#1}}
\usepackage[colorlinks=true, linkcolor=blue, bookmarks=true]{hyperref}

\relax

\usepackage{graphicx}
\usepackage{color}
\usepackage{subfigure}

\newcommand{\ba}{\begin{eqnarray}}
\newcommand{\ea}{\end{eqnarray}}
\newcommand{\be}{\begin{equation}}
\newcommand{\ee}{\end{equation}}

\newcommand{\la}{\lambda}

\newcommand{\ra}{\rightarrow}

\newcommand{\LF}{\left(}
\newcommand{\RF}{\right)}
\newcommand{\LT}{\left[}
\newcommand{\RT}{\right]}

\newcommand{\pd}{\partial}
\newcommand{\D}{\nabla}

\newcommand{\diag}{\mathop{\mathrm{diag}}\nolimits}

\newcommand{\Fc}{\mathcal{F}}
\newcommand{\Hc}{\mathcal{H}}

\newcommand{\Kc}{\mathcal{K}}
\newcommand{\Lc}{\mathcal{L}}

\newcommand{\Wc}{\mathcal{W}}
\newcommand{\Pc}{\mathcal{P}}

\begin{document}

\begin{titlepage}
\begin{flushright}
\end{flushright}
\vfill
\begin{center}
{\Large \bf Cosmological perturbations in non-local \vskip 3mm higher-derivative gravity}

\vskip 10mm

{\large Ben Craps$^{a,b}$, Tim De Jonckheere$^{a,c}$, Alexey S.\ Koshelev$^a$}
\vskip 7mm

$^a$ Theoretische Natuurkunde, Vrije Universiteit Brussel, and \\
\hspace*{0.15cm}  International Solvay Institutes, 
Pleinlaan 2, B-1050 Brussels, Belgium \\
$^b$ Laboratoire de Physique Th\'eorique, Ecole Normale Sup\'erieure,\\ 24 rue Lhomond, F-75231 Paris Cedex 05, France \\
$^c$ Ghent University, Department of Physics and Astronomy,\\ Krijgslaan 281-S9, 9000 Gent, Belgium \\

\vskip 3mm
\vskip 3mm
{\small\noindent  {\tt Ben.Craps@vub.ac.be, Tim.DeJonckheere@UGent.be, alexey.koshelev@vub.ac.be}}

\end{center}
\vfill

\begin{center}
{\bf ABSTRACT}
\vspace{3mm}
\end{center}

We study cosmological perturbations in a non-local higher-derivative model of 
gravity introduced by Biswas, Mazumdar and Siegel. We extend previous work, 
which had focused on classical scalar perturbations around a cosine hyperbolic 
bounce solution, in three ways. First, we point out the existence of a 
Starobinsky solution in this model, which is more attractive from a 
phenomenological point of view (even though it has no bounce). Second, we study 
classical vector and tensor perturbations. Third, we show how to quantize scalar 
and tensor perturbations in a de Sitter phase (for choices of parameters such 
that the model is ghost-free). Our results show that the model is well-behaved 
at this level, and are very similar to corresponding results in local $f(R)$ 
models. {In particular, for the Starobinsky solution of non-local 
higher-derivative gravity, we find the same tensor-to-scalar ratio as for the 
conventional Starobinsky model.}

\end{titlepage}

\tableofcontents


\section{Introduction}

The resolution of cosmological singularities is an outstanding question in 
theoretical cosmology. Inflationary models are often geodesically incomplete 
\cite{Borde:2001nh}, so one needs to understand what happens at (or replaces) 
the singularity in order to explain how the universe emerged in a state that 
enabled inflation. In alternatives to inflation that involve a transition from 
a 
contracting universe into an expanding one, understanding this (possibly 
singular) transition is even more crucial, since it determines how cosmological 
perturbations generated in the contracting phase match onto perturbations in 
the 
expanding phase.

The expectation that an ultraviolet (UV) complete theory of quantum gravity 
should resolve cosmological singularities has motivated the construction of 
many 
string theory models; see for instance 
\cite{Liu:2002yd,Cornalba:2003kd,Durin:2005ix,Craps:2006yb,Gasperini:2007vw,
Berkooz:2007nm,McAllister:2007bg,Das:2009zze, McFadden:2010na,  
Craps:2010bg,Craps:2011sp,Brandenberger:2011et,Burgess:2011fa,Kounnas:2013yda} 
for reviews. Even though interesting mechanisms for singularity resolution have 
emerged, they are often plagued by technical problems or ambiguities. It 
remains 
therefore an open question whether and how string theory resolves cosmological 
singularities. 

An alternative approach is to explore bottom-up models inspired by string 
theory, and a lot of work has happened along these lines. In particular, it is 
known that the UV properties of general relativity can be ameliorated by higher 
derivative corrections. A well-studied class of models is $f(R)$ gravity 
\cite{Mukhanov:1990me,Hwang:2001qk,Sotiriou:2010,Faraoni:2008,Buchdahl:1970}, 
which has the advantage that the equations governing cosmological perturbations 
are tractable. However, models with terms up to a certain order in derivatives 
generically contain ghosts (negative-norm or negative-energy states). This 
problem has been circumvented in a class of higher derivative {\it non-local} 
models of gravity introduced in \cite{Biswas:2005qr}. These models, which we 
will review in Section~\ref{sec:model}, are inspired by infinite-derivative 
structures appearing in string field theory, and should be viewed as effective 
actions (which indeed tend to be non-local). In \cite{Biswas:2005qr}, a 
ghost-free model was constructed and an exact bouncing, spatially flat FLRW 
solution was found.\footnote{While originally it was believed that the model 
constructed in \cite{Biswas:2005qr} is asymptotically-free, the more complete 
analysis in \cite{Biswas:2011ar} showed that it is not, but that it is possible 
to construct an asymptotically free model of gravity by adding additional 
non-local terms to the action. In the present paper, we will focus on the 
original model of \cite{Biswas:2005qr}, leaving the possible generalization of 
our results to asymptotically free models to future work.} This solution 
represents a universe bouncing between contracting and expanding de Sitter 
phases supported by a cosmological constant. Interpreting this cosmological 
constant as a simplified model for the energy density driving inflation, the 
bouncing solution can model the inflationary phase of our universe. Of course 
it 
should really be supplemented by a graceful exit into conventional big bang 
cosmology, but so far this can only be achieved in an approximate way 
\cite{Biswas:2005qr}.

Among the most important predictions of inflationary cosmology are nearly 
scale-invariant spectra of adiabatic density perturbations and gravity waves, so 
it is very 
important to investigate how perturbations behave in the model of 
\cite{Biswas:2005qr}. A first analysis of scalar perturbations was performed in 
\cite{Alexander:2007zm}, 
where the crucial assumption was made that the modification of gravity only 
affects the background evolution, while the perturbations are governed by 
Einstein gravity. Perturbations that only depend on time (and therefore model 
super-Hubble perturbations) were subsequently studied in \cite{Biswas:2010zk}, 
where it was concluded that in the late-time de Sitter phase all modes decay, 
except for one constant mode (which was suggested to be relevant for seeding 
density perturbations). However, the constant mode identified in 
\cite{Biswas:2010zk} is really a gauge artifact (it can be removed by a 
constant 
rescaling of the spatial coordinates). A careful study of all classical scalar 
perturbations, with dependence on both time and space (including sub-Hubble 
modes as well as super-Hubble modes) was undertaken in \cite{Biswas:2012bp}. 
This involved the development of non-trivial techniques to deal with the 
relevant non-local actions. The conclusion of this work is that (at least in 
appropriate parameter ranges) all modes are 
non-singular during the bounce, and decay in the late-time de Sitter phase. 
Since in this simplified model the de Sitter phase is supported by a 
cosmological constant rather than a dynamical inflaton field, this is the 
desired result. At late times, the higher-derivative corrections to general 
relativity are unimportant, so one can expect that including a dynamical 
inflaton will give rise to a nearly scale-invariant spectrum of adiabatic 
density perturbations, as it does in general relativity. 

A more realistic inflationary solution was found in $R+R^2$ gravity, which is a 
{\em local} higher derivative model. It is the
Starobinsky solution \cite{Starobinsky:1982}, which has an inflationary phase 
followed by an exit from inflation, producing a nearly scale invariant spectrum 
of perturbations. We show that this spacetime also emerges as a solution of 
non-local gravity. An important difference with the already mentioned bouncing 
solutions is that in the case of the Starobinsky 
solution inflation is driven by higher derivative terms while in the known 
bouncing solutions inflation is driven by a cosmological constant. 

One question that was not addressed in \cite{Biswas:2012bp} is 
how vector and tensor perturbations behave in these non-local models. We 
present 
a simple argument showing that linearized vector and tensor perturbations are 
governed by {\em local} evolutions equations, which are essentially identical 
to 
those of $f(R)$ gravity models and thus much simpler than those of scalar 
perturbations. 
We compute vector and tensor perturbations in the 
bouncing solution of \cite{Biswas:2005qr}, thereby completing the analysis of 
classical cosmological perturbations in this model. 
We also comment on the growth of anisotropies in a contracting phase.
In addition, we compute the behavior of vector and tensor modes around the 
Starobinsky solution.

Another question we address in the present paper is how to quantize cosmological 
perturbations. Local $f(R)$ gravity models give rise to only one physical mode 
in the 
scalar sector of perturbations. Generically, non-local models with higher 
derivatives give rise to many degrees of freedom. However, by imposing certain 
restrictions on the non-local operator involved, one can ensure that the model 
has only one physical excitation (as we have in 
local $f(R)$ Lagrangians). {We quantize scalar and tensor perturbations in the 
de Sitter regime of the Staribinsky solution and compute their power spectra.} 

The remainder of this paper is organized as follows. In 
Section~\ref{sec:model}, we review the model of \cite{Biswas:2005qr}, 
considering the bouncing solution and presenting the Starobinsky solution in 
this framework.
In Section~\ref{sec:perturbations} we revisit scalar perturbations 
and derive the general equations for vector and tensor perturbations. We apply 
them to the bouncing and Starobinsky backgrounds.
In Section~\ref{sec:quantumPert} we discuss quantum perturbations. We 
conclude with a summary and outlook in Section~\ref{sec:summary}.

\section{The model and background solutions}\label{sec:model}
\setcounter{equation}{0}
\subsection{The model}

The string-inspired higher derivative non-local model of gravity we will 
consider is described by the action \cite{Biswas:2005qr}
\begin{equation}
S=\int 
d^4x\sqrt{-g}\left(\frac{M_P^2}2R+\frac{\lambda}2R\Fc(\Box)R-\Lambda + 
\Lc_M\right).
\label{model}
\end{equation}
Here $M_P$ is the Planck mass, $\Lambda$ is a cosmological constant, $\Lc_M$ is 
a matter Lagrangian\footnote{In this article we will focus on cases in which 
$\Lc_M$ contains at most radiation.} and $\lambda$ is a dimensionless parameter 
measuring the effect of the $O(R^2)$ corrections. We work in four dimensions and 
also limit ourselves to $O(R^2)$ corrections.

The operator $\Fc(\Box)$, which is inspired by structures appearing in string 
field theory, is the ingredient making the difference with other modifications 
of gravity. It is an analytic function with real coefficients of the 
d'Alembertian $\Box=\D^\mu\D_\mu$: 
\begin{equation}
\Fc(\Box)=\sum_{n\geq0}{f_{n}}\Box^n/M_*^{2n}.
\end{equation} 
The new mass scale $M_*$ determines the characteristic scale of the 
modification 
of gravity. Hereafter we absorb $M_*$ in the coefficients $f_n$.

The equations of motion derived from the action (\ref{model}) are
\begin{eqnarray}
[M_P^2+2\lambda\Fc(\Box)R]G^\mu_\nu&=&-\frac{\lambda}{2} R \Fc(\Box) 
R\delta^\mu_{\nu}+2\lambda(\nabla^\mu\pd_\nu-\delta^\mu_{\nu}\Box)\Fc(\Box) 
R\nonumber\\
&&+\lambda{\Kc}^\mu_\nu-\frac{\lambda}{2}\delta^\mu_{\nu}\left({\Kc}
^\sigma_\sigma+\bar\Kc\right)+{T}^\mu_\nu-\Lambda \delta^\mu_{\nu},
\label{EOM}
\end{eqnarray}
where we have defined
\begin{equation}
{\Kc}^\mu_\nu=\sum_{n=1}^\infty{f}_n\sum_{l=0}^{n-1}\pd^\mu  R^{(l)}  \pd_\nu  
R^{(n-l-1)},~\bar\Kc=\sum_{n=1}^\infty{f}_n\sum_{l=0}^{n-1} R^{(l)}    
R^{(n-l)}.
\end{equation}
Here $G^\mu_\nu$ is the Einstein tensor, $R^{(n)}=\Box^nR$  and $T^\mu_\nu$ is 
the matter energy momentum tensor. The trace equation reads
\begin{equation}
-M_P^2R={T}-4\Lambda
-6\lambda\Box\Fc(\Box) R-\lambda({\Kc} +2\bar\Kc),
\label{EOMtrace}
\end{equation}
where quantities without indices denote the trace. This equation is clearly much 
simpler than (\ref{EOM}).

\subsection{Ansatz}\label{sec:ansatz}

Particular background solutions were found by employing the ansatz
\be
\Box R=r_1R+r_2,
\label{ansatz}
\ee
where $r_{1,2}$ are constants. One can then easily compute the $\Box^n R$ 
expression recurrently. Clearly, this simplifies the equations of motion 
considerably but it is apparently non-trivial to satisfy the ansatz itself. 
Substitution of the ansatz in the trace of the Einstein equations 
(\ref{EOMtrace}) 
yields three types of terms which are constant, linear and quadratic in 
curvature, respectively. The following are three relations to cancel each 
structure separately:
\begin{equation}
\Fc^{(1)}(r_1)=0,~\frac{r_2}{r_1}=-\frac{M_P^2-6\lambda\Fc_1r_1}{2\lambda[
\Fc_1-f_0 ] } ,  
~\Lambda=-\frac{r_2M_P^2}{4r_1},
\label{r2lambda}
\end{equation}
where $\Fc_1=\Fc(r_1)$ and $\Fc^{(1)}$ is the first derivative with respect to 
its
argument. Using these expressions, the equations of motion after substituting 
the ansatz are local; we refer to them as Einstein-Schmidt equations:
\ba
\begin{aligned}
2\lambda\Fc_1\LF R + 3r_1\RF G^{\mu}_{\nu} &= T^{\mu}_{\nu} + 2\lambda \Fc_1 
\LT g^{\mu\rho}\nabla_{\rho}\pd_{\nu} R - \frac{1}{4}\delta^{\mu}_{\nu}\LF R^2 
+ 
4r_1 R + r_2\RF\RT.
\end{aligned}
\label{Eseq}
\ea

The same local equations of motion can also be obtained from an $R+R^2$ gravity 
model with action
\ba 
S = \int d^4 x \sqrt{-g}\LF R + \frac{1}{6r_1}R^2 - 2\Lambda_{f(R)} + 
2\Lc_{M,f(R)} \RF,
\label{fRactionWithMatter2}
\ea
where 
\ba
\begin{aligned}
\Lambda_{f(R)} = \frac{-r_2}{2r_1} = \frac{2\Lambda}{M_p^2} && \text{ and } &&  
\Lc_{M,f(R)} = \frac{\Lc_M}{6\lambda\Fc_1 r_1}.
\label{LocalRadvsNonLoc}
\end{aligned}
\ea
We focus here on background solutions of flat FLRW type, so we look for metrics 
of the form
\be
g_{\mu\nu}=\diag(-1,a^2(t),a^2(t),a^2(t)),
\label{FLRW}
\ee
where $a(t)$ is the scale factor. Solutions of the local equations of motion 
that also satisfy the ansatz are solutions of the non-local model. Therefore any 
solution of $R+R^2$ gravity that solves the ansatz is a solution of non-local 
higher derivative gravity~(\ref{model}). Note that after using the ansatz the 
model has no smooth GR limit $\lambda\to 0$ anymore.

\subsection{Bouncing solution}\label{sec:bounce}
One can show~\cite{Biswas:2005qr,Biswas:2012bp} that
\be
a(t)=a_0\cosh(\omega t)
\label{cosh}
\ee
is a solution to the ansatz relation with $r_1=2\omega^2$ and $r_2=-6r_1^2$. 
From e.g.~\cite{Biswas:2005qr} we know that a cosine hyperbolic bounce is a 
solution of $R+R^2$ gravity if some ghostlike (i.e. $\rho_{f(R)}<0$) radiation 
is added. It is therefore also a solution of the non-local model with radiation. 
From analyzing the $00$ component of (\ref{Eseq}) the energy density of the 
traceless radiation in the non-local model is found to be
\begin{equation}
\label{extraradiation}
\rho_0=-\frac{27}{2}\la\Fc_1r_1^2.
\end{equation}
This implies $\Fc_1<0$ in order to avoid ghost components. Comparison of 
$\rho_0$ with $\rho_{0f(R)}$ by making use of (\ref{LocalRadvsNonLoc}) shows 
that positive radiation in the non-local model indeed corresponds to ghostlike 
radiation in the local $R+R^2$ model. 

The cosine hyperbolic bounce is a non-singular solution which reaches a de 
Sitter phase asymptotically at $t\ra \pm \infty$. The bouncing solution is 
therefore said to reach a late time inflating phase. This inflationary phase 
however has no graceful exit and therefore only serves as a toy model.

\subsection{Starobinsky solution}\label{sec:Starobinsky}

Now we present an inflationary solution of the non-local model which at some 
point in time stops inflating. 
The scale factor
\ba
a(t)=a_0\sqrt{t_s - t}e^{\sigma (t_s-t)^2}
\label{StarobinskySol2}
\ea
is an approximate solution of $R+R^2$ gravity without cosmological constant at 
early times when $\left|t\right|\ll t_s$. {It is known as the Starobinsky 
solution. When $\left|t\right|\ll t_s$ and moreover also 
\be\label{StarobinskyCond}
\left|\sigma t_s^ 2 \right|\gg \frac{1}{4},
\ee 
the Starobinsky solution can be approximated by a de Sitter phase. The condition 
(\ref{StarobinskyCond}) guarantees a nearly constant Hubble rate. As is clear 
from (\ref{fRactionWithMatter2}), this corresponds to the regime where the $R^2$ 
correction to gravity dominates. The parameters $\sigma$ and $t_s$ can be 
determined from observations, e.g., from the normalization of the scalar power 
spectrum and from its spectral tilt.}

If a cosmological constant is added, it can be checked that 
(\ref{StarobinskySol2}) becomes an exact solution to the equations of motion of 
$R+R^2$ gravity for all times. It can also be checked that this background 
solves the ansatz (\ref{ansatz}) too. So we conclude that it is also a solution 
of the non-local model. The coefficients for which the ansatz is satisfied, are 
$r_1=-12\sigma$ and $r_2=192\sigma^2= \frac{4}{3}r_1^2$.  {The condition 
(\ref{StarobinskyCond}) can then be expressed as $12 H^2 / r_1 \gg 1$.} Notice 
that exponential expansion is only reached for $\sigma<0$. This means that 
$r_1>0$ and $\Lambda<0$. In this case, the universe with scale factor 
(\ref{StarobinskySol2}) undergoes inflation when $t\ll t_s$, after which it has 
a graceful exit. If one were to consider the scale factor at later times, one 
would find that the universe reaches a maximum size at $t_s - t = 
\sqrt{-1/(4\sigma)}$, after which it contracts due to the negative cosmological 
constant until it eventually hits a singularity at $t=t_s$.

To see whether radiation is needed for having the Starobinsky solution, the $00$ 
component of the Einstein-Schmidt equations (\ref{Eseq}) is checked. It turns 
out that the $00$ equation is satisfied without any radiation. {When the 
Starobinsky model is coupled to particle physics models, radiation should be 
produced during reheating. This goes beyond the scope of the present article, in 
which we will focus on observables that can be computed in the inflationary 
phase.}

\section{Classical perturbations}\label{sec:perturbations}
\setcounter{equation}{0}

In this section we analyze the behavior of classical perturbations in non-local 
higher 
derivative gravity. In particular, we discuss their behavior around the 
cosine hyperbolic bounce and in the Starobinsky background. Scalar 
perturbations 
in the bouncing solution were already discussed in~\cite{Biswas:2012bp}. We 
briefly review their analysis and extend it to vector and tensor perturbations, 
as well as to the Starobinsky solution.

\subsection{Scalar perturbations}\label{scalarmodes}

Scalar perturbations are introduced as
\begin{equation}
ds^2=a(\tau)^2\left[-(1+2\phi)d\tau^2-2\pd_i \beta d\tau 
dx^i+((1-2\psi)\delta_{ij}+2\pd_i\pd_j\gamma)dx^idx^j\right],
\label{mFr}
\end{equation}
where $\tau$ is the conformal time related to the cosmic one as 
$a(\tau)d\tau=dt$.

It is practical to use gauge-invariant  variables to avoid gauge fixing issues. 
These variables are two Bardeen potentials, which fully determine the scalar 
perturbations of the metric tensor 
\cite{Bardeen:1980kt,Mukhanov:1990me,Hwang:2001fb}
\begin{equation}
\Phi=\phi-\frac{1}{a}(a\vartheta)^\prime=\phi-\dot 
\chi,\qquad\Psi=\psi+\Hc\vartheta=\psi+H\chi,
\label{GIvars}
\end{equation}
where  $\chi=a\beta+a^2\dot\gamma$, $\vartheta=\beta+\gamma'$, $\Hc(\tau)=a'/a$, 
and differentiation with respect to the conformal time $\tau$ is denoted by a 
prime.

Before considering these perturbations in non-local higher derivative gravity 
\cite{Biswas:2012bp}, we first briefly summarize what inflationary theories 
typically tell us about scalar metric perturbations.

Standard inflation leads to an exponential expansion of the physical 
wavelengths of fluctuations and thus effectively brings perturbations that are 
initially inside the horizon, to the outside where they get frozen in. Inside 
the horizon, sub-Hubble fluctuations are present due to quantum fluctuations of 
the inflaton field. These sub-Hubble metric fluctuations described by the 
Bardeen 
potential are oscillatory with a constant amplitude. During inflation they are 
carried outside the horizon and once this has happened perturbations are 
frozen in and the Bardeen potential is approximately a constant.

$f(R)$ gravity models provide another way of generating inflation. Because of a 
conformal equivalence between $f(R)$ gravity and normal Einstein gravity with 
addition of scalar field matter, metric perturbations will behave more or less 
in the same way as in standard inflation. This means that scalar perturbations 
are 
also oscillating when they are sub-Hubble but after crossing the horizon 
they get frozen in and become approximately constant on super-Hubble scales.

Now we turn our attention to the behavior in the non-local model. Technically, 
scalar perturbations are subject to very complicated equations and 
the only tractable configurations are those satisfying the ansatz 
(\ref{ansatz}) at the background level. We focus here on two particular 
solutions, namely the bouncing and the Starobinsky solution. Both have a de 
Sitter phase, which is the relevant region we will focus on. Stability under 
scalar perturbations near the bounce was already addressed in 
\cite{Biswas:2012bp} and also the behavior during  a de Sitter phase was 
presented. Here we only present the key results of \cite{Biswas:2012bp} and 
focus on the physical interpretation.  

We immediately write down the two coupled closed equations which read
\ba
\Pc\zeta&=&0,\label{eq341}\\
\frac{\Fc(\Box)-\Fc_1}{\Box_B-r_1}\zeta+\Fc_1[\delta 
R_\text{GI}+(R_B+3r_1)(\Phi-\Psi)]&=&0,
\label{eq342}
\ea
where
\ba
\zeta&=&\delta\Box R_B+(\Box_B-r_1)\delta R_{GI},\label{zeta1}\\
\Pc &\equiv &\LT \pd^{\mu}R_B\pd_{\mu} + 2\LF r_1R_B + r_2\RF\RT 
\frac{\Fc\LF\Box_B\RF - \Fc_1}{\LF\Box_B - r_1\RF^2} + 
3\Fc\LF\Box_B\RF\nonumber\\
&&+ (R_B + 3r_1)\frac{\Fc\LF\Box_B\RF - \Fc_1}{\Box_B - r_1},
\label{TraceEqPert2}\\
\Box_B 
&=&-\frac{1}{a^2}\pd_\tau^2-2\frac{a'}{a^3}\pd_\tau-\frac{k^2}{a^2}
=-\pd_t^2-3\frac{\dot a}{a}\pd_t-\frac{k^2}{a^2}\\
\delta\Box&=&\frac{1}{a^2}\left[2\Phi\left(\pd_\tau^2+2\frac{a'}{a}
\pd_\tau\right)+\left(\Phi'+3\Psi'\right)\pd_\tau\right],\\
\delta\!R_{GI}&=& 
6\Box_B\Psi-2R_B\Phi-6\frac{a'}{a^3}(\Phi'+\Psi')+2\frac{k^2}{a^2} 
(\Phi+\Psi).
\ea
Here the first equation is the perturbation of the trace equation and the 
second comes from considering the $ij$ component of the Einstein equations when 
$i\neq j$.
Essentially, $\zeta$ is the variation of the ansatz (\ref{ansatz}). This ansatz 
is a crucial element in finding background solutions of the non-local model. 
However, being an ansatz it only holds at the background level and will in 
general be perturbed. Here $\delta R_{GI}$ is a gauge 
invariant quantity defined as $\delta R_{GI}=\delta R-R_B'(\beta+\gamma')$.  
If $\zeta=0$, then 
the ansatz is unperturbed and the equations of motion are local as 
in $R+R^2$ gravity even at the perturbed level. Therefore we say that $\zeta$ 
denotes whether perturbations are influenced by non-localities or not. $k$ is 
the length of the comoving spatial momentum.

For the full picture two more equations can be derived which relate 
metric perturbations to energy density and velocity perturbations of the fluid. 
They are the $0i$ and $00$ component of the Einstein equations and were 
computed in \cite{Biswas:2012bp}. Both in GR and in $R+R^2$ gravity the 
perturbed trace equation would be identically zero. Moreover in GR the $ij$ 
component reduces 
to $\Phi=\Psi$. The subscript $B$ designates background quantities. 

We are mainly interested here in the behavior during a de Sitter phase because 
this mimics an inflationary situation. As derived in \cite{Biswas:2012bp} the 
variation of the trace under scalar perturbations can then be written as
\ba
\LF\Box_B - r_1 \RF\Wc\LF\Box_B\RF\delta R_{GI}= 0,
\label{TraceDeSitter2}
\ea
where $\Wc\LF\Box\RF$ is defined as
\ba
\Wc\LF\Box\RF \equiv 3\Fc\LF\Box\RF + (R_B+3r_1)\frac{\Fc\LF \Box\RF - 
\Fc_1}{\Box-r_1}.
\label{WBox}
\ea
The solution $\delta R_{GI}$ can be written as a linear superposition of 
eigenmodes of the d'Alembertian operator with corresponding eigenvalues 
$\omega_i^2$ that are roots of $\Wc\LF\omega^2\RF\LF\omega^2 - r_1\RF=0$.
The eigenmodes $\delta R^{(i)}_{GI}$ satisfy 
\ba 
\LF\Box_B - \omega^2_i\RF\delta R^{(i)}_{GI}=0.
\label{EigenEq}
\ea
The most general solution to this equation can be written in terms of the Bessel 
function $J_{\nu}$ and the Neumann function $Y_{\nu}$ as
\ba
\delta R^{(i)}_{GI} = (-k\tau)^{\frac{3}{2}}\LT d_{1k}J_\nu(-k\tau) + 
d_{2k}Y_\nu(-k\tau)\RT
\label{Besselsolution}
\ea
with $d_{1k}$ and $d_{2k}$ constant in time and 
\ba
\nu = \sqrt{\frac{9}{4} - \frac{\omega_i^2}{H^2}}.
\label{nu}
\ea
\subsubsection{Bouncing solution}\label{sec:BouncingScalar}
In case of the cosine hyperbolic bounce, the de Sitter regime is reached at late 
times, 
consider the long wavelength modes. At late times 
when the cosmological constant dominates and causes the exponential expansion of 
the universe. The non-local modification to the action (\ref{model}) has no 
influence and the behavior of perturbations should be the same as in Einstein 
gravity with a cosmological constant. In that case perturbations are decaying. 
Growing modes in this regime would make the late-time expanding phase unstable 
\cite{Biswas:2012bp}. So we demand that during this late-time de Sitter phase 
scalar modes are decaying which is only the case if every root of 
$\Wc\LF\Box\RF$ is such that the real part $\nu_R$ of $\nu$ satisfies
\ba
\left| \nu_R\right| < \frac{3}{2}.
\label{ConstraintNu}
\ea
It can be checked that also the mode $\omega_i^2 =r_1$ which is always present 
and which corresponds to the mode of $R+R^2$ gravity\footnote{This is the mode 
for which the ansatz is unperturbed.}, is decaying.

\subsubsection{Starobinsky solution}\label{sec:StarobScal}
The Starobinsky solution reaches an exponentially expanding phase at early times 
caused by the non-local higher derivative term in the action (\ref{model}). 
By analyzing (\ref{EigenEq}) in the short wavelength limit it is found that, as 
in $R+R^2$ gravity, short-wavelength perturbations are oscillatory and decay as 
$1/a$. Once the perturbations reach the long wavelength regime they should 
freeze out and become approximately constant. The roots of $\Wc\LF\Box\RF$ 
should therefore satisfy 
\ba
\left| \nu_R \right| \approx \frac{3}{2}.
\ea
Extra decaying modes will not influence the phenomenology so the modes for 
which $\left| \nu_R \right| < 3/2$ are also allowed but they are not 
interesting.

The mode corresponding to $\omega_i^ 2=r_1$ also exhibits
$\nu_R\approx 3/2$. This follows from considering the 
Starobinsky solution in the regime when it is approximately de Sitter, taking 
$r_1=-12\sigma$ and applying the corresponding limits as discussed in 
Section~\ref{sec:Starobinsky}. In this regime, we have
\begin{equation} \label{HStarobinsky}
a(t)\approx a_0\sqrt{t_s}e^{-2\sigma t_s t}\text{ and }H\approx -2\sigma t_s.
\end{equation}
This yields for (\ref{nu})
\begin{equation}
\nu \approx \sqrt{\frac{9}{4} + \frac{3}{\sigma t_s^2}}\to \frac32,
\end{equation}
provided that {(\ref{StarobinskyCond}) holds}. 
Taking into account the small-argument asymptotic behaviors  $J_{3/2}(x)\sim 
x^{3/2}$ and 
$Y_{3/2}(x)\sim x^{-3/2}$, we find that the mode $\omega_i^ 2=r_1$ is 
approximately 
a constant as expected.

The two remaining Einstein equations, namely the $00$ and $0i$ components, in 
general provide conditions involving the energy density and velocity 
perturbation of matter. Since the Starobinsky solution is obtained in a model 
without matter or radiation, these equations simply impose additional conditions 
on the Bardeen potentials. Denoting $u=\Fc\LF\Box_B\RF \delta R_{GI}$, the 
remaining Einstein equations in a de Sitter limit are
\begin{equation}
\begin{aligned}
-6\lambda\frac{\Hc^2}{a^2} u + 4\lambda\Fc_1\LF R_B +3r_1\RF\LF 
\frac{k^2}{a^2}\Psi + 3\frac{\Hc}{a^2}\LF\Psi^{'}+\Hc\Phi\RF\RF\\
 + 2\lambda\Box_B u + \lambda R_B u + \frac{2\lambda}{a^2}\LF u^{''}-\Hc 
u^{'}\RF =0,   
\end{aligned}
\label{00DeSitter2}
\end{equation}
\begin{equation}
2\Fc_1\LF R_B + 3r_1\RF\LF\Psi^{'} + \Hc\Phi\RF - u^{'} + \Hc u = 0.
\label{0iDeSitter2}
\end{equation}
Notice that using this notation, the last equation contains at most first order 
derivatives and can therefore serve as a constraint on the initial data for the 
Bardeen potentials.

\subsection{Vector perturbations}\label{sec:vector}

Vector perturbations in GR die out during inflation and 
are 
therefore unimportant. As long 
as there is no additional rotational anisotropy in the matter tensor, 
angular momentum is conserved, which leads to a decaying solution of the 
metric perturbation. The same  holds for $f(R)$ gravity, but need not be true  
 in more general modified gravity.

Vector modes are introduced as 
\begin{equation}
ds^2 = -a^2(\tau)\left[d\tau^2 - 2b_idx^id\tau + (\delta_{ij} 
+\partial_ic_j +\partial_j c_i)dx^idx^j\right].\label{dsvector}
\end{equation}
By definition the vector modes $b_i , c_i$ are transverse, i.e., 
$\partial_ib^{i}=0$ and  $\partial_ic^{i}=0$. These vector modes 
are related through a gauge transformation and only two independent gauge 
invariant degrees of freedom remain, namely $\Psi_i=b_i + c'_i $
\cite{Bardeen:1980kt,Mukhanov:1990me,Hwang:2001qk}.

At the linearized level,  
$\delta R_{GI}=0$ and 
$\delta\Box f(t)=0$ under vector perturbations. That is because $R$ as well as 
$f(t)$ are scalars while vector and scalar perturbations do not mix at
linear order. This leads to
\begin{equation}
\left[M_P^2 + 2\lambda \Fc\left(\Box_B\right)R_B\right]\LF 
-\frac{1}{2a^2}\Delta 
\Psi_i\RF =  \delta T_{i}^{0},
\end{equation}
where $\delta T^{0}_{i} = \left(\rho_B + 
p_B\right)\delta u_i/a$, with $\delta u_i$ being a 
perturbation 
of the fluid 4-velocity. From $\nabla_{\mu}T^{\mu}_{i}=0$, it follows that 
$\left[a^3\left(\rho_B + p_B\right) \delta u_i\right]'=0$. This expresses 
angular 
momentum conservation, and implies that $\delta T^{0}_{i}=L_k / a^4$, 
with $L_k$ is the angular momentum of the rotational mode.
Then in Fourier space 
vector modes can be written as
\begin{equation}
k^2\Psi_i = \frac{2L_k}{a^2\LF M_P^2 + 2\lambda 
\Fc\left(\Box_B\right)R_B\RF},
\end{equation}
which reduces to the GR result in the limit of  $\la \ra 0$. 

Analyzing the behavior of the latter expression on a class of solutions 
satisfying the ansatz (\ref{ansatz}) we find that
\begin{equation}
k^2 \Psi_i = \frac{L_k}{\lambda \Fc_1 a^2\left(R_B + 3r_1\right)}.
\label{vectorevolve}
\end{equation}
A vector mode thus behaves in exactly the same way as in $R+R^2$ gravity. This 
could have been anticipated, because the conditions $\delta R_{GI}=\delta\Box 
R_B=0$ imply that the ansatz is unperturbed. The classically equivalent $R+R^2$ 
model of gravity thus remains valid at the perturbed level and vector modes 
behave as in local higher derivative gravity \cite{Hwang:2001qk}.

\subsubsection{Bouncing solution}\label{sec:BounceVec}

Now we specialize to the bounce solution (\ref{cosh}), and study the behavior 
of $\Psi_i$ in (\ref{vectorevolve}), both near the bounce and in the late-time 
de Sitter phase. We are interested in answering the following two questions: Do 
we recover the GR limit at late times? And do vector perturbation pass the 
bounce smoothly? Below we will show that vector perturbations at late times do 
behave as in GR. Secondly, in order for the bouncing mechanism not to be 
destroyed, perturbations should be finite near the bounce. Since the bounce 
itself is non-singular, this will indeed be the case.

First we focus on the late-time behavior. When $t\gg\omega^{-1}$ one enters the 
de Sitter phase and the modification term in (\ref{vectorevolve}) becomes 
negligible (see also \cite{Biswas:2005qr}). In this regime the scale factor is 
given by $a(t)\approx a_0 e^{\omega t}/2$. From $H(t)= \omega \tanh(\omega t)$ 
it follows that at late times $H$ is approximately a constant, which in turn 
implies that $R_B$ is a constant. So when $t\gg\omega^{-1}$ the behavior of 
vector perturbations is $\Psi_i\propto 1/a^2$, which is exactly the result one 
can find in GR (see e.g. \cite{Hwang:2001qk,Hwang:2001zt}). 

Near the bounce then, $a\rightarrow a_0$ and $H\rightarrow 0$. Expressed in 
cosmic time $R_B=6\left(\dot{H}+2H^2\right)$. Since $H$ is very small near the 
bounce, we neglect the $H^2$ term. Since $H$ is smooth near the bounce, its 
time 
derivative is well defined and finite, and $R_B$ is certainly finite. In fact 
one can easily show that for the cosine hyperbolic bounce $R_B\approx 6\omega 
^2$. Since $R_B$ is decreasing in the contracting phase while it is increasing 
in the expanding phase, we can conclude that a vector mode starting out in the 
contracting phase, grows near the bounce, reaches a maximal size at the bounce 
and then decays again.

\subsubsection{Starobinsky solution}\label{StarobinskyVec}

Around an inflationary solution the analysis is equivalent to the late-time 
limit of the bouncing solution. Approximating the Starobinsky solution as a de 
Sitter one, the constancy of background curvature implies that the vector modes 
(\ref{vectorevolve}) decay like $1/a^2$ in non-local higher derivative gravity. 
In an inflationary context non-local gravity therefore predicts no cosmological 
vector modes.

\subsection{ Tensor perturbations}\label{sec:tensor}

Unlike scalar perturbations, in standard inflation tensor modes do not couple to 
the inflaton field. 
However, the evolution of tensor modes is described by a covariant Klein-Gordon 
equation, which means that inside the horizon they are oscillatory 
solutions. On these small scales one can approximate the metric with a 
local Minkowski metric so that indeed the sub-Hubble tensor modes describe free 
gravitational 
waves. During inflation the tensor modes are carried outside the horizon such 
that 
on super-Hubble scales they get frozen in.

Tensor perturbations $h_{ij}$ are introduced as 
\begin{equation}
ds^2 = -a^2\left(\tau\right) d\tau^2 + a^2\left(\tau\right)\left(\delta_{ij} + 
2h_{ij}\left(\textbf{x},\tau\right)\right)dx^idx^j.\label{dstensor}
\end{equation}
These modes are by definition symmetric, traceless and transverse, i.e. $h^i_i 
= 
0$ and $\partial_i h^i_j=0$, so that one ends up with 2 independent modes 
\cite{Bardeen:1980kt,Mukhanov:1990me,Hwang:2001qk}). Tensor modes are gauge 
invariant.

For the same reason discussed above for vector modes, $\delta R=\delta \Box 
f(t)=0$ under tensor perturbations.
The equation of motion becomes
\begin{equation}
h^{\prime\prime}_{ij} + \LF 2\Hc + \frac{2\lambda\LF\Fc\LF\Box_B\RF R_B 
\RF^{\prime}}{M_P^2 + 2\lambda \Fc\left(\Box_B\right)R_B}\RF  h^{\prime}_{ij} + 
k^2 h_{ij} = 0,
\end{equation}
which again reduces to the GR equation in the limit $\la \ra 0$.
Turning to background solutions satisfying the ansatz (\ref{ansatz}) we have in 
cosmic time
\begin{equation}
 \ddot{h}_{ij}+\left(3H + \frac{\dot{R_B}}{R_B+3r_1}\right)\dot{h}_{ij}+ 
\frac{k^2}{a^2}h_{ij}=0.
 \label{tensorevolcosmotime}
\end{equation}
This is the same evolution as in $R+R^2$ gravity
\cite{Hwang:2001qk}.

\subsubsection{Bouncing solution}\label{sec:bounceTensor}

In the late-time regime of the bouncing solution  (\ref{cosh}), the 
${k^2}/{a^2}$ term in equation (\ref{tensorevolcosmotime}) can be neglected 
because the scale factor becomes exponentially large. Since the scalar 
curvature 
is constant at late times one can drop the $\dot{R}_B/(R_B+3r_1)$ term in 
Eq.~(\ref{tensorevolcosmotime}) as well. With these approximations  
Eq.~(\ref{tensorevolcosmotime}) can easily be integrated
\begin{equation}
h_{ij}(k,t) = c_{ij}(k)  - d_{ij}(k)\int{\frac{dt}{a^3}},
\label{tensorintegralcosmo}
\end{equation}
which is exactly the result one finds in GR \cite{Hwang:2001qk,Hwang:2001zt}.

Near the bounce, we can expand equation (\ref{tensorevolcosmotime}) up to linear 
order in $t$ resulting in 
\begin{equation}
 \ddot{h}_{ij}+4\omega^2 t \dot{h}_{ij}+ \frac{k^2}{a_0^2}h_{ij}=0.
 \label{tensorevolexpand}
\end{equation}
In order to get a picture of how tensor perturbations behave near the bounce, we 
decompose $h=\alpha\kappa$ and write $\frac{d}{dt}\ln\left[a^3\left(R_B + 
3r_1\right)\right]=v$. Then (\ref{tensorevolcosmotime}) becomes
\begin{equation}\frac{\ddot{\alpha}}{\alpha} + 
\frac{\ddot{\kappa}}{\kappa} 
+\frac{\dot{\alpha}}{\alpha}\left(v+2\frac{\dot{\kappa}}{\kappa}\right) + 
v\frac{\dot{\kappa}}{\kappa}  + \frac{k^2}{a^2} = 0.
\end{equation}
We fix $\kappa$ by the requirement
\begin{equation}
v+2\frac{\dot{\kappa}}{\kappa}=0~\Rightarrow~\kappa (t) = e^{-\frac{1}{2}\int{v 
dt}} = \frac{1}{\sqrt{a^3\left(R_B + 3r_1\right)}}
\end{equation}
The resulting $\kappa$ is regular around the bounce. Thanks to the choice of 
$\kappa$ we are left with the second order differential equation on $\alpha$ 
only
\begin{equation}
\ddot{\alpha} + \left(\frac{k^2}{a^2} - \frac{\dot{v}}{2} - 
\frac{v^2}{4}\right)\alpha = 0,
\label{alphaevol}
\end{equation}
which is moreover a Schr\"odinger type equation with zero energy and potential 
\begin{equation}
V = \frac{\dot{v}}{2} + \frac{v^2}{4}-\frac{k^2}{a^2}.
\end{equation}
Near the bounce $v=4\omega^2 t$ so that the potential becomes just $V = 
(2a_0^2\omega^2 - k^2)/a_0^2$. The evolution equation (\ref{alphaevol}) 
for $\alpha$ becomes a (possibly inverted) harmonic oscillator for which the 
evolution is described by
\begin{equation}
\alpha(t) = Ae^{\sqrt{V}t} + Be^{-\sqrt{V}t}.
\end{equation}
where $A,B$ are integration constants. For small wavenumbers 
$k^2<2a_0^2\omega^2$, we get a linear combination of an exponentially growing 
and a decaying mode. They both pass the bounce point at $t=0$ smoothly. Tensor 
perturbations with shorter wavelengths for which $k^2 > 2a_0^2\omega^2$ become 
oscillatory near the bounce. This condition on oscillatory solutions only holds 
near the bounce and has nothing to do with the definition of sub-Hubble 
perturbations. Indeed, at zeroth order the Hubble radius $H^{-1}$ near the 
bounce is infinite such that all perturbations are inside the Hubble radius. We 
can conclude that a tensor mode generated in the contracting phase will start 
oscillating near the bounce, pass the bounce smoothly and become constant at 
late times. 

\subsubsection{Starobinsky solution}\label{sec:StarobTensor}
Again the analysis of tensor modes around the Starobinsky background is 
comparable to the late-time analysis around the bouncing solution. However, the 
idea now is that perturbations start out sub-Hubble at the beginning of 
inflation. The variation of curvature is negligible during a quasi de Sitter 
stage and inside the Hubble radius the friction term in 
(\ref{tensorevolcosmotime}) can be neglected. In conformal time the evolution 
equation reduces to that of a harmonic oscillator.
Once the perturbations reach the long wavelength regime $k\ll Ha$ the friction 
term dominates and (\ref{tensorevolcosmotime}) is solved by 
(\ref{tensorintegralcosmo}). As in the late time regime of the cosine hyperbolic 
background, tensor modes freeze out. Like standard inflation or $R+R^2$ gravity, 
non-local higher derivative gravity predicts the generation of cosmological 
gravitational waves.

\subsection{Special case: homogeneous anisotropic perturbations}\label{sec:beta}

A special class of homogeneous diagonal perturbations of the metric, 
is introduced as $ds^2 = -dt^2 + a^2e^{2\eta_i}dx_{i}^2$ with 
$\sum_i\eta_i=0$. 
This is a metric of the type Bianchi~I which describes an anisotropic universe 
and 
is a test-bed for the analysis of anisotropic perturbations. In GR the $\eta_i$
appear as a new effective matter component in the Einstein equations energy 
whose energy  scales as $a^{-6}$, leading to problems in a contracting phase 
(see for instance \cite{Cai:2013vm}). In the context of a non-singular bounce, 
this growth of anisotropies leads to a fine tuning problem on the initial 
perturbations in the contracting phase. We now study this issue in the context 
of non-local gravity.

In the perturbative regime ($\eta_i \ll 1$), the metric 
becomes 
\begin{equation}
ds^2 = -dt^2 + a^2(1+{2\eta_i})dx_{i}^2,\label{dsbeta}
\end{equation}
and one can try to identify the $\eta_i$ within the general formalism of linear 
perturbations introduced before.
The question is whether they 
belong to the scalar, vector or tensor sector. However it should be noted that 
although one can always split a $4\times 4$  matrix into a spin 0, spin 1 and a 
spin 2 piece, the explicit distinction between pure scalar modes, vector and 
tensor modes is ill-defined if $k= 0$ because of the trivially vanishing 
spatial 
derivatives. Still, one can treat the $\eta_i$ modes as if they were 
inhomogeneous, go to Fourier space, and then take the limit $k\rightarrow 0$. 
Since the $\eta_i$ modes are diagonal and traceless, it follows that there are 
two independent modes. One of them will contain a longitudinal part while the 
other one will be completely transverse. If the $\eta_i$ modes are written in 
vector notation as $\vec{\eta}= (\eta_1, \eta_2, \eta_3)$, then the 
tracelessness condition causes the most general mode $\vec{\eta}$ to be a linear 
combination of a $(1,1,-2)$ and a $(1,-1,0)$ mode. These correspond  
respectively to a scalar and 
a tensor mode. From the form of the line element we see that all $\eta_i$ 
should satisfy the same evolution equation. Both the scalar and the tensor mode 
should then also satisfy this same equation in the limit $k\rightarrow 0$.  

To derive the equation in question we note that comparing the metric 
(\ref{dsbeta}) with the general perturbed line element (the combination of 
(\ref{mFr}, \ref{dsvector}, \ref{dstensor})) it follows that the only non-zero 
modes are $\gamma$ and $h_{ii}$. If $\eta_i$ is a scalar mode, it can be written 
as $\eta_i = \partial^2_i \gamma$ and  from $\sum_i{\eta_i}=0$ it follows that 
$\triangle \gamma=0$ or in Fourier space $k^2 = 0$. It is the only non-zero 
scalar mode so the gauge invariant potentials can be written as $\Phi = 
-\frac{1}{2}\mathcal{H} \gamma' - \frac{1}{2}\gamma^{''}$ and $\Psi =\frac{1}{2} 
\mathcal{H}\gamma'$. 
A tensor mode in general does not perturb the ansatz, so neither should the 
particular scalar modes we are interested in (since they should satisfy the 
same 
equation). This leads to huge simplifications in (\ref{eq341}) and 
(\ref{eq342}) 
for scalar modes by setting $\zeta=0$. The equations reduce to 
\begin{equation}
\delta R_{GI} + \left(R_B + 3r_{1}\right)\left(\Phi - \Psi\right) =0.
\end{equation}
Substituting $\Phi$ and $\Psi$ in terms of the scalar mode $\gamma$ and dividing 
by the factor
$-\left(R_B + 3r_1\right)/2$, we get
\begin{equation}
\ddot{\gamma} + \left(3H + \frac{\dot{R}_B}{R_B + 3r_1}\right)\dot{\gamma}=0,
\label{scalarlimitk0}
\end{equation}
which is exactly the same as the evolution equation for tensor modes in the 
limit of $k\rightarrow 0$. We can conclude that homogeneous (i.e. $k\rightarrow 
0$) and traceless perturbations on the metric diagonal  consist of a scalar and 
a tensor mode and their evolution is described by (\ref{scalarlimitk0}). Thus 
all work on homogeneous perturbations of the form (\ref{dsbeta}) fits into our 
more general framework, which uses a decomposition into scalars, vectors and 
tensors.
Integrating (\ref{scalarlimitk0}), we can estimate the behavior of  $\sigma^2 
\equiv \sum_i \dot{\eta}_i^2$:
\ba
\sigma^2 \propto \frac{1}{a^6\LF R_B + 3r_1\RF^2}.
\ea
In a bouncing context, scalar curvature is decreasing in the contracting phase 
such that anisotropies grow at least as $a^{-6}$. Bounce solutions in non-local 
gravity thus also suffer from a fine tuning problem (but not an infinite one, 
since there is a finite minimal scale factor).

\section{Quantum perturbations}\label{sec:quantumPert}
\setcounter{equation}{0}

In standard inflation, cosmological perturbations have a quantum mechanical 
origin. {At the beginning of inflation, modes of observational interest have 
wavelengths small compared to the Hubble radius. The standard assumption is that 
such modes start out in the Bunch-Davies vacuum.\footnote{{We do not consider 
possible complications for trans-Planckian modes; see, for instance, 
\cite{Martin:2001,Kaloper:2002}, for some discussion on this.}} During 
inflation, their wavelengths become larger than the Hubble radius. Computation 
of the subsequent evolution is facilitated by the fact that certain variables 
remain constant while they stay outside the horizon.} Typically, inflationary 
theories predict (nearly) scale invariant power spectra of scalar and tensor 
modes after inflation. In this section, we discuss how perturbations can be 
quantized in non-local higher derivative gravity, at least in the de Sitter 
regime of the Starobinsky solution (which is our solution of most 
phenomenological interest). {Restricting to the de Sitter regime amounts to 
focusing on physical wavelengths that satisfy
\be\label{wavelength}
\frac{k}{a}\ll \frac{1}{\sqrt{r_1}}.
\ee
Indeed, as shown in section~7 of \cite{Mukhanov:1990me}, such modes behave as if 
they propagated in de Sitter space.}
We calculate the power spectra of scalars and tensors {as well as the} 
tensor-to-scalar ratio.

Since we consider linear perturbations, we construct the second variation of the 
action, first in a 
covariant way. The metric is perturbed as
$g_{\mu\nu}=g_{B\mu\nu}+h_{\mu\nu}$. We decompose $S=S_0 + S_1$ where
\ba 
\begin{aligned}
S_0 &= \int d^4x\ \sqrt{-g}\left[ \frac{M_P^2}2 R -\Lambda\right],~
S_1&=\int d^4 x \sqrt{-g}\frac\lambda2 R
\Fc(\Box)R.
\end{aligned}
\ea
There is a  problem computing $\delta^2 S_1$ around an inflationary background 
with a graceful 
exit because $(\delta\Box) R_B\neq 0$, such that the non-local function 
$\Fc$ 
itself is varied and we have no simple way of expressing the second variation 
of 
the action in terms of scalar modes. {We are able to avoid this problem by 
focusing on modes that exit the horizon during the de Sitter phase of the 
Starobinsky solution, during which $R_B$ is approximately constant} and 
$R_{B\mu\nu}=R_Bg_{\mu\nu}/4$.

For $\delta^2 S_0$ we use the results of \cite{Christensen:1980}. 
The second variation is
\ba
\begin{aligned}
\delta^2 S_{0}=&\int d^4x\sqrt{-g}\frac{M_P^2}2\bigg[\frac14 h_{\mu\nu}\Box_B
h^{\mu\nu}-\frac14h\Box_B h+\frac12h\nabla_{B\mu}\nabla_{B\rho}
h^{\mu\rho}+\bigg.\\
&\frac12\nabla_{B\mu} h^{\mu\rho}\nabla_{B\nu}
h^\nu_\rho-\bigg.\frac1{48}R_B(h^2+2h^\mu_\nu h^\nu_\mu)\bigg]\equiv\int
d^4x\sqrt{-g}\frac{M_P^2}2\delta_0.
\end{aligned}\label{d2properlambda0Rconst}
\ea
The variation of the non-local higher derivative action $S_1$ can now be easily
expressed in terms of $\delta R_{GI}$ and $\delta_0$,
\ba
\begin{aligned}
\delta^2 S_{1}=&\int
dx^4\sqrt{-g}\left(\lambda \Fc\LF\Box_B\RF R_B\delta_0+\frac\lambda2\delta 
R_{GI}\Fc(\Box_B)\delta R_{GI}\right),
\end{aligned}\label{d2proper1Rconst}
\ea
{where we have neglected terms containing the variation of the 
d'Alembertian operator because they are negligible compared to the second term 
of (\ref{d2proper1Rconst}).}\footnote{{To see this, one uses 
equations (3.4) and (3.2) of \cite{Biswas:2012bp}, combined with the fact that 
$\delta\Box R_B$ is negligible compared to $\LF\Box_B - r_1\RF\delta R_{GI}$. 
The latter statement can be verified by expressing both quantities in terms of 
the Bardeen potentials $\Phi$ and $\Psi$, and using (\ref{StarobinskyCond}), 
(\ref{wavelength}), $\dot{H}/H^2\ll 1$ and $\dot{R}_B / H^3 \ll 1$.}}
Since the de Sitter background satisfies the ansatz (\ref{ansatz}), the total 
action quadratic in metric perturbations reduces to 
\begin{equation}
\begin{split}
\delta^2 S=&\int
dx^4\sqrt{-g}\frac\lambda2\left(2\Fc_1 \LF R_B + 3r_1\RF\delta_0+\delta 
R_{GI}\Fc(\Box_B)\delta R_{GI}\right),
\end{split}\label{d2proper1Rconst2}
\end{equation}

\subsection{Quantizing scalar modes}

We consider scalar perturbations and work in the Newtonian gauge,
\ba
ds^2 = -a^2\LT 1 + 2\Phi\RT d\tau^2 + a^2\LT 1- 2\Psi\RT dx^i dx_i.
\label{newtongauge}
\ea
Although the line element is written in terms of the gauge invariant Bardeen 
potentials, we have to be aware that the gauge is really fixed, namely that 
$\Phi= \phi$ and $\Psi=\psi$. In the Newtonian gauge the variation $\delta_0$ 
is equal to
\ba
\delta_0 = -\frac{1}{a^2}\LT 4k ^2\Psi^2 + 4k^2 \Psi\Phi + 12\Hc\Phi\Psi^{'} + 
6\Hc^2\Phi^2\RT - 6\Psi \Box \Psi.
\label{delta0Newtongauge}
\ea

The Einstein equations lead to a further reduction of the action. They can be 
categorized into two types: dynamical equations and constraint equations 
\cite{Arnowitt:1962,ADM:1962}. The latter contain at 
most first order time derivatives of the perturbed variables. They relate the 
Bardeen potentials to each other and determine constraints on the initial 
hypersurface on which quantisation proceeds. As done in \cite{Mukhanov:1990me} 
for a local higher derivative model, we use here the 0i equation
(\ref{0iDeSitter2}) as a constraint. After a lengthy computation, the second 
variation of the action then reduces to  
\ba
\delta^2 S = \int d^ 4 x \sqrt{-g} \frac{\lambda}{2\Fc_1R_B}u 
\mathcal{W}\LF\Box_B\RF{\LF\Box_B - 
r_1\RF}\frac{1}{\Fc\LF\Box_B\RF} u,
\ea
where total derivatives have been dropped {and where we have 
taken into account that in the inflationary phase of the 
Starobinsky solution $r_1$ is negligible compared to $R_B$ (to see this, use 
(\ref{HStarobinsky})).} Varying with respect to $u$ reproduces the perturbed 
trace equation.

Notice that we have silently assumed that $\Fc\LF\Box\RF$ is invertible, which 
is only the case if $\Fc$ has no roots. Moreover, the non-local theory should 
be ghostfree even at the perturbed level. This means that also $\Wc$ must not 
have any roots. Indeed, a Weierstrass decomposition of $\Wc / \Fc$ into its 
roots shows that every root introduces a classical degree of freedom. As soon as 
more than one root is present, the theory suffers from ghosts. The only root 
that is always present is the one which does not perturb the ansatz, so the one 
which is also present in $R+R^2$ gravity. The action can then be written as
\ba
\delta^2 S = \frac{1}{2}\int d^ 4 x \sqrt{-g} ue^{\gamma\LF\Box_B\RF}\Box_B u,
\ea
where 
\be
e^{\gamma\LF\Box_B\RF}=\frac{\lambda\mathcal{W}
\LF\Box_B\RF}{\Fc_1R_B\Fc\LF\Box_B\RF}
\ee
and $\gamma\LF\Box_B\RF$ is 
an entire function. A simple redefinition
\ba
\tilde{v}=e^{\gamma\LF\Box_B\RF / 2} u
\label{utv}
\ea
turns the action into a local one
\ba
\delta^2 S = \frac{1}{2}\int d^ 4 x \sqrt{-g} \tilde{v}{\LF\Box_B 
- r_1\RF} 
\tilde{v}.
\ea
This local action in a de Sitter background can be easily written as
\ba
\delta^2 S = \frac{1}{2}\int d^4 x \LT v^{'2} + v\Delta v +{\LF 
2\Hc^2 - a^2 r_1\RF} v^2\RT,
\label{SecondVarV_i}
\ea
where 
\ba \label{vav}
v = a\tilde{v}.
\ea 
We retrieve the action of a scalar particle with time-dependent mass in 
Minkowski 
space, which can be straightforwardly quantized.

We want to compute the power spectrum 
$\left|\delta_{\Phi}(\vec{k},\tau)\right|^2$, which is determined by the 
equal-time two point correlation function of the Bardeen potential $\Phi$  by
\ba
\bra{0}\hat{\Phi}(\vec{x},\tau)\hat{\Phi}(\vec{x}+\vec{r},\tau)\ket{0} = 
\int\limits_{0}^{+\infty} \frac{dk}{k}\frac{\sin(kr)}{kr} 
\left|\delta_{\Phi}(\vec{k},\tau)\right|^2.
\label{PowerSpectrumPhi}
\ea 
Here, the de Sitter invariant vacuum $\ket{0}$ is obtained by selecting 
negative frequency modes via the initial 
conditions 
\ba
\begin{aligned} 
v(\vec{k},\tau_0) &= \frac{1}{k^{3/2}}\LF \Hc_0 + ik\RF e^{ik\tau_0},\\
v^{'}(\vec{k},\tau_0) &= \frac{i}{k^{1/2}}\LF \Hc_0 + ik - 
\frac{i\Hc^{'}_0}{k}\RF e^{ik\tau_0},
\label{InitialCond}
\end{aligned}
\ea
where $\Hc_0$ is the Hubble constant at an early time $\tau_0$.

First of all we need to relate the two-point correlator of $v$ to that of the 
Bardeen 
potential. The Einstein equations in de Sitter space allow us to relate $\Phi$, 
$\Psi$ and $u$. Combining (\ref{0iDeSitter2}) with (\ref{00DeSitter2}) leads to 
the constraints
\ba
\Psi = \frac{1}{2\Fc_1R_B} u,~
\Phi = -\Psi,
\label{DeSitterPsiPhi}
\ea
where the last equality has been obtained by also using the $ij$ component. 
Using (\ref{utv}) and (\ref{vav}), we find that
\ba
\begin{aligned}
\bra{0}\hat{\Phi}(\vec{x},\tau)\hat{\Phi}(\vec{x}+\vec{r},\tau)\ket{0} =&
\frac{1}{(2\Fc_1R_B)^2}\bra{0}e^{-\gamma\LF\Box_B\RF 
/2}\LF\frac{1}{a}\hat{v}(\vec{x},\tau)\RF\times \\ & e^{-\gamma\LF\Box_B\RF 
/2}\LF\frac{1}{a}\hat{v}(\vec{x}+\vec{r},\tau)\RF\ket{0}\\
=&
\frac{1}{4\pi^2}\frac{1}{(2\Fc_1R_B)^2} \int\limits_{0}^{+\infty} 
\frac{dk}{k}\frac{\sin(kr)}{kr}
k^3\left|e^{-\gamma\LF\Box_B\RF /2}\tilde{v}(\vec{k},\tau)\right|^2.\nonumber
\end{aligned}
\ea
Here $\hat v$ is the quantum field representing the canonically quantized 
classical field $v$.
The vacuum state is defined at some initial time $\tau_0$ corresponding to the 
beginning of inflation such that all initial perturbations are sub-Hubble.
In 
the inflationary context, like the Starobinsky background,
the solution for $v$ implies that 
$\tilde{v}(k,\tau)$ is an eigenstate of the d'Alembertian with 
{eigenvalue $r_1$}. 
It is given by (\ref{Besselsolution}) for 
{$\nu\approx 3/2$}.
For such a mode, the initial conditions hold 
not only at time $\tau_0$ but at any time {in the inflationary 
regime (i.e. as long as $r_1\ll R_B$)} because it is {an 
approximate} solution to the 
equation of motion for $v$ (which can be easily derived from the action 
(\ref{SecondVarV_i}) after writing everything in Fourier {space}).

The primordial power spectrum of the 
sub-Hubble modes becomes
\ba
\left|\delta_{\Phi}(\vec{k},\tau)\right|^2 \approx \frac{k^2}{16\pi^2 
a^2}\frac{1}{\lambda\Fc_1R_B}\frac13,
\ea
where we re-expressed {$e^{-\gamma(r_1)}$} through $\Fc$ and $\Wc$ and used 
(\ref{WBox}) to obtain\footnote{{The fact that the second term in (\ref{WBox}) 
is zero when $\Wc$ is evaluated at $r_1$ easily follows from (\ref{r2lambda}).}} 
 $\Fc_1/\Wc(r_1)=1/3$.
Like in a local higher derivative theory, the primordial spectrum depends on 
the 
physical wavelength as $\left|\delta_\Phi\right|^2\propto 1/\lambda_{ph}^2$ 
where $\lambda_{ph}\sim a/k$. Since (\ref{InitialCond})  holds at any time for 
a nearly massless mode, the spectrum in the long wavelength limit can be found 
by just taking the limit $k\ll \Hc$ of (\ref{InitialCond}). In that way we 
retrieve the following scale invariant power spectrum:
\ba
\left|\delta_{\Phi}(\vec{k},\tau)\right|^2 \approx 
\frac{H^2}{16\pi^2}\frac{1}{\lambda\Fc_1R_B}\frac13.
\ea
As a consistency check we note that our results for the power spectra reproduce 
the well known answers of $R+R^2$ gravity in a de Sitter 
limit~\cite{Mukhanov:1990me}. It is so since we have reduced our consideration 
to the regime of the pure de Sitter inflation when the effects of 
non-localities vanish. 

{Usually in $R+R^2$ gravity one does not calculate the spectrum of $\Phi$ but 
rather that of 
\ba 
\mathcal{R} \equiv \Psi + \frac{H}{\dot{R}_B}\delta R_{GI}.
\label{defCurlyR}
\ea 
During the de Sitter phase of the Starobinsky solution, we can make use of 
(\ref{DeSitterPsiPhi}), fill in that $u=\Fc\LF\Box\RF\delta R_{GI}$ and remember 
that $\delta R_{GI}$ satisfies (\ref{EigenEq}) with eigenvalue $r_1$ and 
$\dot{H}\ll H^2$ such that $\mathcal{R}$ reduces to
\ba
\mathcal{R} \approx -\frac{H^2}{\dot{H}}\Phi.
\label{DeSitterCurlyR}
\ea
The advantage of calculating the spectrum of $\mathcal{R}$ is that during 
a slow-roll inflation this quantity is constant in the super-Hubble limit. 
During the Hubble radius crossing we thus find a scale invariant primordial 
spectrum
\ba
\left|\delta_{\mathcal{R}}(\vec{k},\tau)\right|^2 \approx 
\frac{H_{k=Ha}^4}{16\pi^2\dot{H}^2_{k=Ha}}\frac{1}{\lambda\Fc_1R_B}\frac13.
\ea
}

\subsection{Quantizing tensor modes}\label{sec:QuantTensor}

Since tensor perturbations behave in a more local way than scalar 
perturbations, 
we expect it should be possible to quantize them around more general spacetimes 
than de Sitter. However, in order to be able to compare power spectra of 
tensors 
and scalars, we will restrict our attention to de Sitter.

Using the fact that $\delta R_{GI}=0$ under tensor perturbations, the second 
variation (\ref{d2proper1Rconst2}) of the action under tensor perturbations 
reads
\ba
\delta^2 S = \int
d^4x\sqrt{-g}\lambda\Fc_1 \xi\left(\frac{1}{4}h_{\mu\nu}\Box_B h^{\mu\nu} - 
\frac{1}{24}R_B h^{\mu}_{\nu}h^{\nu}_{\mu}\RF,
\label{SecondVarTensors2}
\ea
where $\xi=R_B+3r_1$, which in the de Sitter phase of the Starobinsky solution 
reduces to $\xi\approx R_B$. Tensors only perturb the spatial $3\times 3$ part 
of the 
metric such that we can write $h_{\mu\nu}=a^2 h_{ij}\delta_{i\mu}\delta_{j\nu}$ 
and $h^{\mu\nu}= h_{ij}\delta_i^\mu\delta_j^\nu/a^ 2$. Furthermore, in a flat 
FLRW universe, $\sqrt{-g}=a^4$, from which it follows that 
\ba
\delta^2 S = \int d^4x \frac{\lambda\Fc_1 \xi a^2}{4}\LF h^{'2}_{ij} + 
h_{ij}\Delta h_{ij}\RF.
\label{SecondVarTensors4}
\ea
This is indeed the second variation of the action of $R+R^2$ gravity (with a 
cosmological constant) under tensor perturbations in a de Sitter background. In 
\cite{Mukhanov:1990me} the power spectrum of tensor perturbations in $R+R^2$ 
gravity was already calculated during a de Sitter stage. The quantization 
procedure is completely analogous to the scalar case, but with the difference 
that the action can now be written in such a way that we retrieve a massless 
mode in Minkowski space which describes a gravitational wave 
(see~\cite{Mukhanov:1990me}). 

The primordial power spectrum of sub-Hubble tensor modes in non-local higher 
derivative gravity is thus
\ba
\left|\delta_{h}\right|^2 =\frac{1}{\pi^2 2\lambda \Fc_1\xi}\frac{k^2}{a^2}.
\ea
In the super-Hubble regime, a scale invariant power spectrum is found:
\ba
\left|\delta_{h}\right|^2 =\frac{H^2}{\pi^2 2\lambda \Fc_1\xi}.
\ea 
So, during a de Sitter stage, non-local higher derivative gravity predicts the
same power spectrum as in $R+R^2$ gravity. Essentially this is due to the fact 
that tensors do not perturb the ansatz and therefore non-localities do not play 
a role.

By dividing the tensor and scalar power spectra, taking into account 
$\xi\approx R_B$ {and including a factor of 2 because of the 
graviton  polarizations, we retrieve 
\ba
r = \frac{2\left|\delta_{h}\right|^2}{\left|\delta_{\mathcal R}\right|^2} 
\approx 48 \frac{\dot{H}_{k=Ha}^2}{H_{k=Ha}^4}
\ea}
for the tensor-to-scalar ratio $r$ during the de Sitter phase of the 
Starobinsky 
solution. {In terms of the slow-roll parameter $\epsilon_1 = -\LF\dot{H} / 
H^2\RF_{k=Ha}$ the ratio becomes
\ba 
r = 48\epsilon_1^2.
\ea
The slow-roll parameter is easily  related to the number of $e$-foldings 
$N\equiv \int_{t_i}^{t_f} H dt \approx 1/(2\epsilon_1)$ during the inflationary 
phase~\cite{Felice:2010}. We now retrieve the familiar result
\ba 
r = \frac{12}{N^2}.
\label{tensorScalarEfolding}
\ea
Non-local higher derivative gravity thus predicts the same tensor-to-scalar 
ratio as $R+R^2$ gravity~\cite{Felice:2010}.}

\section{Summary and Outlook}\label{sec:summary}
\setcounter{equation}{0}
The original motivation for studying a class of non-local string field theory 
inspired generalizations of gravity (\ref{model}), arose from the possibility 
of 
constructing a simple bouncing solution for the scale factor of an FLRW 
universe 
without introducing ghosts.  In this paper we have shown the existence of an 
inflationary solution. This is the Starobinsky solution, which was originally 
found in $R+R^2$ gravity. Moreover, by noticing the classical equivalence with 
an $R+R^2$ model of gravity we have shown that every solution of $R+R^2$ 
gravity 
that is also a solution of a particular ansatz (\ref{ansatz}), solves the 
equations of motion of the non-local model. (This point has recently also been 
made in~\cite{Vernov:2014}.)
We have considered vector and tensor perturbations in non-local higher 
derivative gravity. Thanks to the fact that the action is built from scalar 
quantities only, we have managed to show that the equations of motion are 
perturbed under the vector and tensor perturbations in close analogy to $f(R)$ 
gravity models. The crucial technical consequence is that vector and tensor 
perturbations obey local equations of motion leaving all the non-localities at 
the background level. For special solutions that satisfy the simplifying ansatz 
(\ref{ansatz}), vector and tensor modes completely reduce to ones from $f(R)$ 
theories. We have illustrated this by considering their behavior around both a 
bouncing background and the Starobinsky solution. Both solutions have a de 
Sitter phase, enabling us to compare the behavior of these modes with that in 
standard inflationary theories.
As a consequence of the equivalence with perturbations in $f(R)$ gravity, 
vector modes die out with time in an exponentially expanding phase. Non-local 
gravity therefore predicts no cosmological vector modes. In a bouncing context, 
tensor modes start growing in the contracting phase, pass the bounce smoothly 
and get frozen in at late times of the expanding phase. In the vicinity of 
bounce both vector and tensor modes grow, but not without bound. In an 
inflationary context, tensor modes are constant to leading order. In accordance 
with other inflationary models, non-local higher derivative gravity thus 
predicts the existence of cosmological gravitational waves.

Primordial metric perturbations are typically quantum mechanical fluctuations 
generated during inflation. In this paper we have shown how to quantize both 
scalar and tensor perturbations {around the de Sitter phase}. In 
this 
approximation, we have found that the power spectra for scalar and tensor modes 
during inflation agree with those in $R+R^2$ gravity.

A natural direction for future research in this framework is to consider the 
more general model  introduced in \cite{Biswas:2011ar,Koshelev:2013ida}, the 
action of which contains not only the Ricci scalar, but also the Ricci tensor. 
In this model, gravity can be asymptotically free for a certain class of 
parameters. The model allows the same cosine hyperbolic solution, found using a 
generalized ansatz involving also the Ricci tensor. We expect that in this model 
the equations for vector and tensor modes will not reduce to those in $f(R)$ 
gravity, and that one will have to face non-local equations for vector and 
tensor perturbations. 


\section*{Acknowledgments}

We would like to thank K.\ Van Acoleyen and S.\ Vernov for useful 
conversations. This work was supported in part by the Belgian Federal Science 
Policy Office through the Interuniversity Attraction Pole P7/37, by 
FWO-Vlaanderen through project G011410N, and by the Vrije Universiteit Brussel 
through the Strategic Research Program ``High-Energy Physics''. A.K.\ is 
supported by an ``FWO-Vlaanderen'' postdoctoral fellowship and also in part by 
RFBR grant 11-01-00894.


\end{document}